\documentclass[grl]{agu2001}
 \usepackage{graphicx}
\authorrunninghead{WENZLER ET AL.}
\titlerunninghead{Reconstructed and measured TSI between 1978 and 2003}
\authoraddr{T. Wenzler,
Max-Planck-Institut f\"ur Sonnensystemforschung, 37191 Katlenburg-Lindau, Germany.
(twenzler@hsz-t.ch)}

\begin{document}

%% ------------------------------------------------------------------------ %%
%
%  ENABLE IMAGE DISPLAY WHILE USING DRAFT MODE
%
%% ------------------------------------------------------------------------ %%
%
% Uncomment the following code (as well as \usepackage{graphicx} above)
% if you need to include images in draft mode
%\setkeys{Gin}{draft=false}
%
% PLEASE NOTE: WHEN YOU SUBMIT YOUR LATEX FILE TO GEMS, COMMENT OUT ANY COMMANDS
% THAT INCLUDE GRAPHICS.
% (See FIGURES section near the end of the file)
%

%% ------------------------------------------------------------------------ %%
%
%  TITLE
%
%% ------------------------------------------------------------------------ %%

\title{Reconstructed and measured total solar irradiance: Is there a secular trend between 1978 and 2003?}

%% ------------------------------------------------------------------------ %%
%
%  AUTHORS AND AFFILIATIONS - 3 methods
%
%% ------------------------------------------------------------------------ %%

% Method 2 (for all journals, except Reviews of Geophysics, which
% should use method 3):
% For more than three author/affiliation blocks,
% use \author{\altaffilmark{}} and \altaffiltext{}
% \altaffilmark will produce footnote;
% matching altaffiltext will appear at bottom of page.
% May use \\ to start a new line.

 \authors{T. Wenzler, \altaffilmark{1,2}
 S. K. Solanki, \altaffilmark{1,3}
 and N. A. Krivova \altaffilmark{1}}

 \altaffiltext{1}
 {Max-Planck-Institut f\"ur Sonnensystemforschung, D-37191 Katlenburg-Lindau, Germany.}

 \altaffiltext{2}{Hochschule f\"ur Technik Z\"urich, CH-8004 Z\"urich,
   Switzerland.}

 \altaffiltext{3}{School of Space Research, Kyung Hee University,
   Yongin, Gyeonggi 446-701, Korea.}

%% ------------------------------------------------------------------------ %%
%
%  ABSTRACT
%
%% ------------------------------------------------------------------------ %%

% Do NOT include any \begin...\end commands within
% the body of the abstract.

\begin{abstract}
Total solar irradiance reconstructed between 1978 and 2003 using solar
surface magnetic field distributions is compared with three composites
of total solar irradiance measurements. A good correspondence is found
with the total solar irradiance composite from PMOD/WRC, with no bias
between the three cycles. The agreement with the other composites (the
ACRIM composite, mainly based on the Active Cavity Radiometer Irradiance 
Monitors I, II \& III, and the IRMB composite from the Institut Royal
Meteorologique Belgique) is significantly poorer. In
particular, a secular increase
in the irradiance exhibited by these composites is not present in the
reconstructions. Hence any secular trend in total solar irradiance between
1978 and 2003 is not due to magnetic fields at the solar surface.
\end{abstract}

%% ------------------------------------------------------------------------ %%
%
%  BEGIN ARTICLE
%
%% ------------------------------------------------------------------------ %%

% The body of the article must start with a \begin{article} command,
% and an \end{article} command must be placed at the end of the file,
% before \end{document}.
%
% If using draft mode \end{article} must follow the references section.

\begin{article}

%% ------------------------------------------------------------------------ %%
%
%  TEXT
%
%% ------------------------------------------------------------------------ %%

\section{Introduction}\label{sec:intro}
Regular monitoring of total solar irradiance (TSI) by space-based radiometers started in 1978 and has continued without major interruptions since then, although no single instrument so far managed to survive longer than a single solar activity cycle. Crosscalibrating and combining the data from the different radiometers into one record is not without problems, and three independent groups have constructed distinct composites of the Sun's total irradiance from the measurements \citep{froehlich2000,froehlich2003b,froehlich2006,willson1997,willsonandmordinov2003,dewitteetal2004}. These time series show substantial differences, in particular regarding the long-term trend of the irradiance. This is best seen as the difference between the TSI during the minima preceding solar cycles 22 and 23 \citep{froehlich2006}.

Here we compare a reconstruction of TSI based on magnetograms and
model atmospheres with the irradiance composites. Since such
reconstructions have turned out to be highly successful in reproducing
the irradiance on the solar cycle time scale
\citep{krivovaetal2003,wenzleretal2006} this comparison provides an
independent test of the composites and in particular of whether any
secular trends in the irradiance can be due to solar surface magnetic
fields. The precent paper does not address the lower 
  irradiance in the minimum between cycles 23 and 24
  \citep{lockwoodandfroehlich2008} since the instruments that recorded
  the data on which the present reconstructions are based stopped operating in 2003.

%\begin{figure*}
%\noindent\includegraphics[width=39pc]{comp_dev_v41.eps}
% \caption{3-month running means of the PMOD (solid line), ACRIM (dotted line) and IRMB (long dashed line) composite records of total solar irradiance between 1978 November 16 and 2004 June 10 (\emph{top panel}). The other three panels show differences between the various composite records. The grey lines are daily values and the black lines are their 1-month running means.}
% \label{fig:compdev}
% \end{figure*}  (henceforth referred to as NSO-SPM)

\section{Data and model}\label{sec:data}
\subsection{Composite records of total solar irradiance measurements}\label{sec:composites}
We consider here three different composites, which are compiled from multiple, cross-calibrated independent radiometric measurements since 1978 November 16. We use the newest PMOD composite of TSI (version 41) from the Physikalisch-Meteorologisches Observatorium Davos/World Radiation Center (PMOD/WRC) in Switzerland \citep{froehlich2000,froehlich2003b,froehlich2006}, the ACRIM composite \citep{willson1997,willsonandmordinov2003} and the IRMB composite from the Institut Royal Meteorologique Belgique \citep[][S. Dewitte priv. comm.]{dewitteetal2004}. %The 3-month running means of the PMOD (solid line), ACRIM (dotted line) and IRMB (long dashed line) composite records of total solar irradiance between November 1978 and April 2004 are presented in the top panel of Fig.~\ref{fig:compdev}. 
The three composite time series are quite different in their longer term trends. %To better illustrate this disparity, we show in the lower panels of Fig.~\ref{fig:compdev} daily differences (grey lines) and their 1-month running means (black lines) between all composite records. 
The PMOD composite has a negligible trend between the two solar minima of 1986 and 1996 (cycles 22 and 23) and has lower TSI values at the three solar maxima of the cycles 21, 22 and 23 than the other two composites \citep{froehlichlean2004,froehlich2003a,willsonandmordinov2003,dewitteetal2004}. The ACRIM composite shows the largest difference between the 2 minima. It is important to note that the differences between the composites not only change gradually, but also display significant jumps.

\subsection{Magnetograms and continuum images}\label{sec:datamagint}
The TSI model employs sets of full-disk magnetograms in Fe\,{\sc i}~8688~${\rm \AA}$ and continuum images obtained daily at the Kitt Peak Vacuum Tower \citep[KPVT;][]{livingstonetal1976a} of the National Solar Observatory (NSO). Data recorded by the NASA/NSO spectromagnetograph (henceforth referred to as NSO-SPM) between 1992 November 21 and 2003 September 21 and by the NASA/NSO 512-channel Diode Array Magnetograph (henceforth referred to as NSO-512) between 1978 November 16 and 1992 April 4 were used for irradiance reconstructions on a total of 3528 days. The NSO-512 magnetograms need to be divided by a factor $f$ in order to make them compatible with the NSO-SPM data. The standard factor recommended by NSO is $f=1.46$. However \citet{wenzleretal2006} have shown that different cross-calibration techniques give different factors between 1.38 and 1.63. A detailed describtion of the data and their treatment is given by \citet{wenzleretal2004,wenzleretal2006}.

\subsection{Model}\label{sec:model}
We use the SATIRE model \citep[Spectral And Total Irradiance
  REconstructions,][]{solankietal2005} described by
\citet{fliggeetal2000a,fliggeetal2000b}, \citet{krivovaetal2003} and
\citet{wenzleretal2004,wenzleretal2005,wenzleretal2006}. It is based
on the assumption that all irradiance changes on time scales of a day
and longer are caused by the evolving distribution of the magnetic field on the solar surface. It makes use of two elements: full disk data (magnetograms for identifying bright magnetic features and continuum images for sunspots and pores) and model atmospheres for each of these components. The model has a single free parameter, $B_{\rm sat}$, the value of the magnetogram signal at which magnetic field in the solar photosphere fills the whole magnetogram pixel. $B_{\rm sat}$ is varied until the best representation of the full considered TSI composite is obtained. We have carried out two sets of TSI reconstructions, those using the standard $f$ value of 1.46, called standard reconstructions (discussed in Sect.~\ref{sec:compstandard}) and those allowing $f$ to vary within the allowed range of 1.38 and 1.63, called optimized reconstructions (discussed in Sect.~\ref{sec:compoptimal}). A detailed describtion of the TSI model used here is given by \citet{wenzleretal2006}.

The length of the reconstructed time series is determined by the
length of time for which the Kitt Peak magnetograms are available. For
this study we decided against combining these reconstruction with 
those from MDI on SoHO, in order to maintain homogeneity in the reconstructed
time series. A combination with MDI would have entailed employing
conversion factors between Kitt Peak magnetograms and MDI magnetograms
\citep[e.g.][]{wenzleretal2004}, which would have introduced further
uncertainty in the long-term trend.

\section{TSI reconstructions in cycles 21, 22 and~23}\label{sec:rec}
We now compare the reconstructed TSI with the three TSI composite records: PMOD, ACRIM and IRMB.

\subsection{Standard reconstruction}\label{sec:compstandard}
Here we compare the standard reconstructions of TSI
\citep{wenzleretal2006} with all TSI composites. This is illustrated
in Fig.~\ref{fig:one}, which in the top panel displays the
reconstructed TSI, while the lower panels show the difference between
measured and reconstructed TSI for all three composites. The best
agreement is achieved with the PMOD composite (second panel). Not only
is the scatter larger in the two lowermost panels, but also the
slopes of the regression lines are noticeably larger, indicating a
significant difference between the long-term trends of these
composites and the reconstruction. 

The corresponding correlation
coefficients, $r_{\rm c}$, between the reconstructed and measured
irradiance, the slopes and $\chi^2$ of the linear regressions as well
as the $B_{\rm sat}$ values are listed in Table~\ref{tab:one} under
``Standard Reconstructions''. For all three composites we find that
the correlation coefficient is lower for the NSO-512 period than for
the NSO-SPM period, probably due to the lower quality of the SPM-512
magnetograms, but also of the early TSI data. The correlation coefficients obtained when comparing the reconstructions with the ACRIM and IRMB composites are always significantly lower than those for the PMOD compilation. This is true for all intervals of time considered, but is most striking for the full period.

%..., both the PMOD composite and the reconstructed irradiance show no trend significantly different from zero \citep{froehlichlean2004,ich2003a}\citep{froehlichlean2004,froehlich2003a}

The ACRIM composite displays a difference of roughly $0.8\:{\rm
  W/m^2}$ between the minima of cycles 22 and 23
\citep{willsonandmordinov2003}, while \citet{dewitteetal2004} obtain a
smaller difference of $0.15 \pm 0.35\:{\rm W/m^2}$. On the other hand,
both the PMOD composite and the reconstructed irradiance show no trend
significantly different from zero
\citep{froehlichlean2004,wenzleretal2006}. The main reason for the
differences during the period before ACRIM-I starts is that in the
PMOD composite \citet{froehlich2006} corrects the data of the Hickey-Frieden (HF) radiometer
on NIMBUS-7 for changes he attributes to an early increase, degradation and other long-term
changes. Next, all three lower panels of Fig.~\ref{fig:one} show a
jump in 1984 which coincides with the repair of the Solar Maximum
Mission (SMM) Spacecraft and the re-start of the ACRIM I data. We
cannot rule out problems in the NSO-512 magnetograms as the cause \citep[see][for
  details]{wenzleretal2006}, but the timing suggests that it could at
least partly be due to a change of
ACRIM I after its extended switch-off during the repair of SMM. 

Another obvious difference between the PMOD and the other composites
is seen during the gap between ACRIM I and II. It is manifested as a very
steep drop at the end of September 1989. This is most likely due to a slip in HF
which was already detected by \citet{leeetal1995} by comparison with
their Earth Radiation Budget Satellite (ERBS) data and later by
\citet{chapmanetal1996} relative to their model. In the IRMB composite 
the tracing of ACRIM II to ACRIM I is performed 
with ERBS data, which do not have the slip in September 1989. This is
responsible for the fact that the ACRIM I and II differences to the
model lie at the same level (close to zero) in the composite. The data
during the ACRIM gap, however, are from HF and so obviously show that
step.

Also, between 1992 and 2003 with the more reliable NSO-SPM data the
reconstructed TSI values agree best with the PMOD composite (second
panel of Fig.~\ref{fig:one} and Table~\ref{tab:one}). During this
period the ACRIM composite is generally higher than the reconstructed
TSI. \citet{froehlich2004a} has removed steps in the TSI from the PMOD
composite in July 1992 and in March 1998, that, he argued, were
produced when the radiometers A and B of ACRIM II were switched.
No such compensation was made in the other composites. After 1996, 
when the VIRGO data started, IRMB uses a different
method to determine the degradation corrections of DIARAD than
underlies the PMOD composite.

\subsection{TSI reconstruction with optimized factor between NSO-512 and NSO-SPM}\label{sec:compoptimal}
Next, we repeated the analysis, but now allowed $f$ to vary within the allowed range of 1.38 and 1.63 in order to optimise the reconstructions to the different composites. 

To best reproduce the PMOD composite irradiance time series
(i.e. achieve the highest $r_{\rm c}$ and lowest $\chi^2$), we need a
slightly higher $f$ of 1.6 compared with the official
factor of 1.46 (see column 3 of Table~\ref{tab:one}), whereas for the
ACRIM and IRMB composites a significantly higher factor of 2.0 is
required, which lies well outside the allowed range
(1.38-1.63). Within this allowed range we receive the best possible
reproduction of the ACRIM and IRMB composites using a factor of
1.63. The correlation coefficients etc.\ between the reconstructions
and ACRIM and IRMB composites are given in the lower half of
Table~\ref{tab:one}. Note that the correlation coefficients for the
ACRIM and IRMB composites are significantly lower in all of the three
considered periods (even for $f=2.0$) than for the PMOD time series.

\section{Conclusions}\label{sec:concl}
\begin{figure*}
 \noindent\includegraphics[width=39pc]{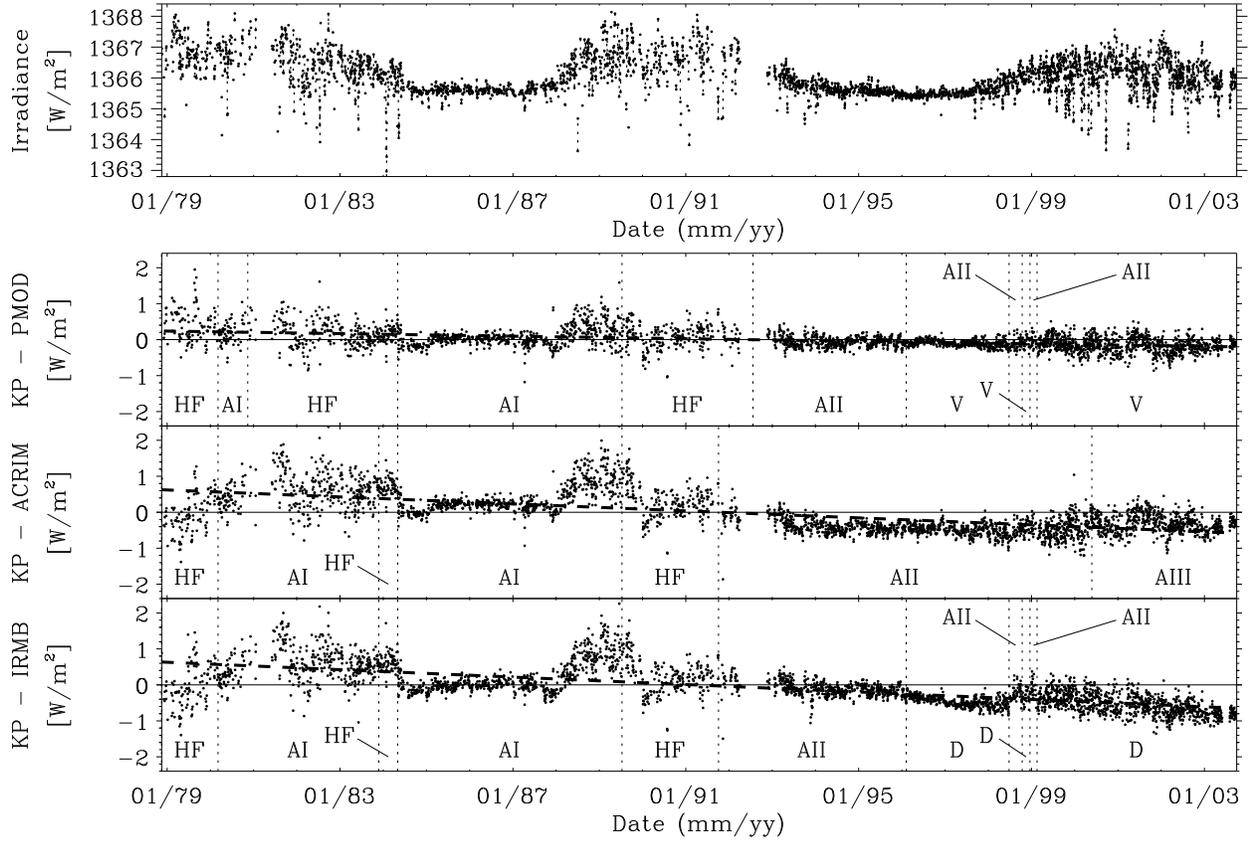}
 \caption{The \emph{top panel} shows the reconstructed daily total solar irradiance based on both NSO-512 and NSO-SPM data for 3528 days between 1978 and 2003, i.e. from the rising phase of cycle 21 to the declining phase of cycle 23. The three \emph{lower panels} show the difference between the reconstructed TSI and the PMOD, ACRIM and IRMB composites, respectively. Each dot represents a daily value. The horizontal solid lines indicate difference~=~0, the thick dashed lines are linear regressions. The dotted vertical lines indicate periods, when the individual data sets from HF, AI, AII \& AIII (ACRIM I, II \& III), V (VIRGO: Variability of solar IRradiance and Gravity Oscillations experiment on SoHO) and D (DIARAD: DIfferential Absolute RADiometer) were used for the composites.}
\label{fig:one}
 \end{figure*}

We have compared the TSI reconstructed between 1978 and 2003 with
three different composites of measured TSI (PMOD, ACRIM and IRMB). Our
model, based on the assumption that the solar irradiance changes are
entirely caused by the evolution of the solar surface magnetic fields,
suggests similar levels of the solar irradiance during the three
minima of the solar cycles 21, 22 and 23 \citep{wenzleretal2006}. With
a single free parameter fixed for the whole period, the model
reproduces the observed irradiance variations in all three cycles 21,
22 and 23 as represented by the PMOD composite. However, the
reconstructed irradiance displays large deviations from the other two
composites (ACRIM and IRMB), as can be judged from
  Fig.~\ref{fig:one} and Table~\ref{tab:one}. This implies that any
secular trend in the irradiance between 1978 and 2003 (e.g., found in the ACRIM and
IRMB composites) cannot be due to the surface magnetic field as sampled by ground-based magnetograms. Such magnetograms do
  miss very small-scale fields in the quiet Sun \citep[see][for reviews]{solanki2009,dewijnetal2009}. The  influence of these weak fields on irradiance, if any, still has to be etsablished.

% TWO-COLUMN figure
%
%

% TWO-COLUMN table
%
\begin{table*}
\centering
\caption{$B_{\rm sat}$ [in G] of the reconstructed TSI, correlation coefficients between reconstructed and composite TSI, $r_\mathrm{c}$, slopes and $\chi^2$ of the linear fits to the scatter plots of modelled total solar irradiance based on NSO-512 and NSO-SPM data vs.\ PMOD, ACRIM and IRMB composite measurements for different periods. The reconstructed TSI values are calculated with different $f$ factors in the upper and lower halves of the table.}
\begin{tabular}{cc@{ -- }cccccc}
\tableline
{Comp.} & \multicolumn{2}{c}{Years} & Magnetogram & $B_{\rm sat}$ & $r_\mathrm{c}$ & Slope & $\chi^2$  \\
\tableline
& \multicolumn{2}{c}{} &  Standard Reconstructions (Sect.~\ref{sec:compstandard})\\
\tableline
PMOD & 1978 & 2003 & NSO-512/1.46 and NSO-SPM & 340 & 0.89 & $1.003 \pm 0.009$ & 0.077 \\
PMOD & 1978 & 1992 & NSO-512/1.46 & 340 & 0.88 & $1.023 \pm 0.015$ & 0.111 \\
PMOD & 1992 & 2003 & NSO-SPM & 340 & 0.93 & $0.937 \pm 0.008$ & 0.031 \\
\tableline
ACRIM & 1978 & 2003 & NSO-512/1.46 and NSO-SPM & 270 & 0.71 & $0.782 \pm 0.013$ & 0.277 \\
ACRIM & 1978 & 1992 & NSO-512/1.46 & 270 & 0.82 & $0.868 \pm 0.017$ & 0.247 \\
ACRIM & 1992 & 2003 & NSO-SPM & 270 & 0.91 & $0.981 \pm 0.010$ & 0.057 \\
\tableline
IRMB & 1978 & 2003 & NSO-512/1.46 and NSO-SPM & 260 & 0.74 & $0.798 \pm 0.012$ & 0.269 \\
IRMB & 1978 & 1992 & NSO-512/1.46 & 260 & 0.81 & $0.943 \pm 0.018$ & 0.268 \\
IRMB & 1992 & 2003 & NSO-SPM & 260 & 0.90 & $0.812 \pm 0.009$ & 0.072 \\
\tableline
& \multicolumn{2}{c}{} &  Optimized Reconstructions (Sect.~\ref{sec:compoptimal})\\
\tableline
PMOD & 1978 & 2003 & NSO-512/1.6 and NSO-SPM & 320 & 0.91 & $0.978 \pm 0.008$ & 0.056 \\
PMOD & 1978 & 1992 & NSO-512/1.6 & 320 & 0.89 & $0.968 \pm  0.014$ & 0.091 \\
PMOD & 1992 & 2003 & NSO-SPM & 320 & 0.94 & $0.983 \pm 0.008$ & 0.032 \\
\tableline
ACRIM & 1978 & 2003 & NSO-512/1.63 and NSO-SPM & 250 & 0.79 & $0.815 \pm 0.011$ & 0.189 \\
ACRIM & 1978 & 1992 & NSO-512/1.63 & 250 & 0.82 & $0.810 \pm 0.015$ & 0.205 \\
ACRIM & 1992 & 2003 & NSO-SPM & 250 & 0.91 & $1.041 \pm 0.011$ & 0.067 \\
\tableline
IRMB & 1978 & 2003 & NSO-512/1.63 and NSO-SPM & 250 & 0.82 & $0.805 \pm 0.010$ & 0.166 \\
IRMB & 1978 & 1992 & NSO-512/1.63 & 250 & 0.82 & $0.858 \pm 0.016$ & 0.204 \\
IRMB & 1992 & 2003 & NSO-SPM & 250 & 0.90 & $0.842 \pm 0.009$ & 0.075 \\
\tableline
\end{tabular}
\label{tab:one}
\end{table*}
%

%% ------------------------------------------------------------------------ %%
%
%  ACKNOWLEDGMENTS
%
%% ------------------------------------------------------------------------ %%

\begin{acknowledgments}
We thank M.~Fligge, D.M.~Fluri and C.~Fr\"ohlich for helpful
discussions. This work was partly supported by the \emph{Deut\-sche
  For\-schungs\-ge\-mein\-schaft, DFG\/} project number SO~711/1-1,
and partly by the WCU grant (No. R31-10016) from the Korean Ministry
of Education, Science and Technology).
\end{acknowledgments}

\end{article}

\end{document}